\documentclass[preprint]{aastex}

\usepackage{rotating}
\usepackage{natbib}

\shorttitle{Modeling Nodal Superhumps}
\shortauthors{Montgomery}

\begin{document}

\title{Numerical Simulations of Naturally Tilted, Retrogradely Precessing, Nodal Superhumping  Accretion Disks}
\author{M.M. Montgomery\altaffilmark{1} }
\affil{$^{1}$Department of Physics, University of Central Florida, Orlando, FL  32816, USA}

\begin{abstract}
Accretion disks around black hole, neutron star, and white dwarf systems are thought to sometimes tilt, retrogradely precess, and produce hump-shaped modulations in light curves that have a period shorter than the orbital period.  Although artificially rotating numerically simulated accretion disks out of the orbital plane and around the line of nodes generates these short-period superhumps and retrograde precession of the disk, no numerical code to date has been shown to produce a disk tilt naturally.   In this work, we report the first naturally tilted disk in non-magnetic Cataclysmic Variables (CVs) using 3D Smoothed Particle Hydrodynamics (SPH).  Our simulations show that after many hundreds of orbital periods, the disk has tilted on its own and this disk tilt is without the aid of radiation sources or magnetic fields.  As the system orbits, the accretion stream strikes the bright spot (which is on the rim of the tilted disk) and flows over and under the disk on different flow paths.  These different flow paths suggest the lift force as a source to disk tilt.  Our results confirm the disk shape, disk structure, and negative superhump period and support the source to disk tilt, source to retrograde precession, and location associated with X-ray and He II emission from the disk as suggested in previous works.  Our results identify the fundamental negative superhump frequency as indicator of disk tilt around the line of nodes.   
\end{abstract}


\section{Introduction}
More than twenty non-magnetic Cataclysmic Variable (CV) Dwarf Novae (DN) systems are known to exhibit negative superhumps in their light curves, the period of which is shorter than the orbital period of the system.  Negative superhumps are suggested to be a consequence of a partially tilted disk (Patterson et al. 1993) or are related to a warped accretion disk (Petterson 1977; Murray \& Armitage 1998; Terquem and Papaloizou 2000; Murray et al. 2002; Foulkes et al. 2006).  Harvey et al. (1995) suggest a retrograde precession of the line of nodes in an accretion disk that is tilted with respect to the orbital plane could yield this type of superhump in its light curve (see also Bonnet-Bidaud, Motch, \& Mouchet 1985 and Barrett, O'Donoghue, \& Warner 1988).  

Pringle (1996) finds an existing warp in an accretion disk can be further modified by axisymmetric radiation pressure forces. Also, an initially flat disk is unstable to warping by a central radiation source. Simulations by Murray \& Armitage (1998) find negligible disk tilt induced by a small vertical resonance (less than one degree, but nonetheless a small tilt).  Using 3D Smoothed Particle Hydrodynamic (SPH) simulations, Wood et al. 2000 (see also Montgomery 2004, 2009a) show that negative superhumps can be artificially induced in light curves when the accretion disk is artificially rotated around the line of nodes and out of the orbital plane.  They also find that a tilted axisymmetric disk retrogradely precesses.  In 3D grid magneto-hydrodynamics (MHD) numerical simulations by Bisikalo et al. (2003, 2004), the disk remains co-planar with the orbital system but a 3D spiral density wave in inner annuli retrogradely precesses due to tidal torques on this vertical wave (Montgomery \& Bisikalo, 2010).  To date, though, no numerical code has been shown to produce a tilted, retrogradely precessing disk that produces negative superhumps in the artificial light curves without invoking magnetic fields, a significant central radiation source, or artificial disk tilt.

In this work, we show that the 3D SPH code by Simpson (1995) does produce naturally tilted disks in non-magnetic CV DN systems without the aid of magnetic fields or radiation sources.

\section{3D SPH Numerical Code}
The 3D SPH codes used and described in Wood et al. (2000), Montgomery (2004, 2009a), Wood et al. (2009), and Wood et al. (2011) have their roots in Simpson (1995), a code that applies to non-magnetic CV systems.  These later versions of the root code have more particles (i.e., more than 100,000), modified particle shapes, varying injection rates, among other changes.  However, none of these changes have resulted in publications of naturally tilted, retrogradely precessing disks and negative superhumps in light curves.  As shown in these works, to induce retrograde precession and negative superhumps, the run is stopped, the disk is artificially rotated, and the run is restarted. Although negative superhumps and retrograde precession are now present, the act to induce the tilt is artificial.  Hence we return to the root code to see if accretion disks can tilt naturally  and, if so, to more easily identify the source of disk tilt. 

The root code (Simpson 1995) uses the Lagrangian method of SPH, has a maximum of 25,000 particles, assumes an ideal gas law with a low adiabatic gamma, does not include radiation effects or magnetic fields.  The only unknown is alpha in the Shukura-Sunyaev (1973) $\alpha$-disk model.  If a disk tilts using this root code, then the source to this disk tilt cannot be due to radiation effects, magnetic fields, and/or MHD.  Although the root code has low resolution, we seek a gross disk tilt and high resolution should not be needed.  In this work, we assume a primary mass $M_{1}$=0.6$M_{\odot}$ and a primary-to-secondary mass ratio $q$=0.45.  The center-to-center distance between stars is $d=0.97 R_{\odot}$ and the orbital period of this system is 2.86 hours.  We assume typical values for viscosity coefficients, $\alpha$=0.5 and $\beta$=0.5, and for smoothing length $h=0.02$.  

\section{Results and Analysis}
The root code (Simpson 1995) does produce a natural disk tilt around the line of nodes but only after a time of several hundred orbital periods has passed.  During the first couple hundred orbital periods, the gas stream strikes the bright spot.  As the disk shape above and below the bright spot is similar, the gas stream passes nearly equally over and under the disk.  After a time of several hundred orbital periods has passed, we start to see changes: a) The disk has settled, forming a cold dense midplane.  b) The disk rim shape is dissimilar above and below the midplane; it resembles a modern day airfoil.  c) The disk has tilted a few degrees around the line of nodes. d) When not at the nodes, the gas stream strikes the disk either slightly above or below the bright spot.  e) The accretion stream and some dislodged bright spot particles pass over and/or under the disk rim.  f) Over approximately one orbital period, the accretion stream (and dislodged bright spot particles) pass mostly over the disk rim for approximately the first half an orbital period and then mostly under the disk rim for approximately the last half an orbital period.  g) The accretion stream splashes onto middle disk annuli of each disk face.  

Figure 1 shows an artificial bolometric light curve (top two panels) and an associated Fourier transform (bottom panel).  To obtain the fractional amplitude bolometric luminosity, the original bolometric light curve (not shown) is divided by a boxcar smoothed light curve, width 203, and then one is subtracted.  The Fourier transform includes orbital periods 700-1200, fifty of which are shown in the middle panel.  The negative superhump modulation is difficult to discern in the the upper two panels, unlike in Figure 6 in Montgomery (2009a) and in Figure 8 in Wood et al. (2009).  The strong signal-to-noise in the lower panel confirms periodicity.  The negative superhump period suggests a tilted disk (see Wood et al., 2000).  The low amplitude of the signal suggests a low disk tilt (see Montgomery  2009a).  

As shown in the third panel, our simulation produces a 2.106 orbit$^{-1}$ frequency, which is twice the negative superhump fundamental frequency, because the luminosity is bolometric:  Due to an optically thick disk, the observer sees emission from one disk face per orbit and thus only one modulation per orbit (i.e., $\nu_{-}$=1.053 orbits$^{-1}$).  Our simulation produces a negative superhump period of $P_{-}$=0.950$P_{orb}$ which agrees with that found in artificially tilted disk simulations (see Table 1 of Montgomery 2009a) and is relative to that found in Table 1 in Wood et al. (2009), a simulation which assumes a different mass-radius relation.  Although not shown, we find for the q=0.35 and q=0.4 simulations $2\nu_{-}$=2.097orbit$^{-1}$ and $2\nu_{-}$=2.103 orbit$^{-1}$ or $P_{-}$=0.951$P_{orb}$ and $P_{-}$=0.950$P_{orb}$, respectively.  Similarly, these values agree with those found from artificially rotated disks as well as those found from a differentially rotating disk with an attached ring generated from the gas stream splashing onto the disk (see Montgomery 2009a).  Of note is the similar negative superhump periods produced in naturally tilted disk simulations.  If naturally tilted disks produce similar negative superhump periods, then observed negative superhump periods may be due to other effects affecting the disk simultaneously such as positive superhumps.  

Figure 2 shows snapshot edge-on and face-on views of the accretion stream and disk at time 1230.0$P_{orb}$ in the simulation.  The lower panel shows an isotropically scaled, edge-on view whereas the upper left panel's aspect ratio has been changed to emphasize the disk tilt and accretion stream paths.  The line of nodes is nearly perpendicular to the page in the edge-on views.  As only a disk tilt is seen, only a fundamental frequency (i.e., no harmonics) should be present in the Fourier transform.  We identify this fundamental negative superhump frequency as the indicator of disk tilt around the line of nodes. 

In both the face-on view (upper right panel) and edge-on views of Figure 2, black represents both the coolest and warmest bolometric emission.  The coolest emission is from the accretion stream and the from rim of the disk.  The warmest emission is from the innermost annuli of face-on disk.  Only the coolest emission is seen in the edge-on views.  Like the artificially rotated disks in Montgomery (2009a), the face-on view shows coolest emission from the outermost disk annulus, colored black, with inner annuli increasing in bolometric luminosity per particle.  However, unlike the artificially rotated disks in Montgomery (2009a), the geometric thickness of the naturally tilted disk is not constant:  The outermost disk annulus (black) is geometrically thin whereas the next two adjacent inner annuli (red and yellow, respectively) are thicker.  Bottom line is that the disk structure is different in naturally tilted disks:  Enough time has passed for the coldest particles to sink toward the disk midplane.  Now, a thin, cold, dense, tilted midplane has formed.

Also shown in Figure 2, some disk particles involved in the collision at the bright spot (orange crosses) are dislodged and join the accretion stream flowing under the disk (black crosses). Because these dislodged particles have more energy, they travel further before splashing onto the disk.  Some of the higher energy particles in this path (orange crosses) appear to reach the rim of the disk at (x,y)=(-0.6,-0.3) whereas the cooler particles (black crosses) reach the outer half of the orange disk annuli.  Because the accretion stream lands on middle and outer annuli, where emission is less per particle and more particles are found in the disk, the negative superhump should not be prominent in the light curve.  In other words, we should not expect to see strong negative superhumps in the light curves because the accretion stream does not land on the innermost annuli of the disk where the highest emission per particle is generated.   

Figure 3 shows a cartoon of one random orbit of a tilted-disk.  Cartoon is based on the simulations.  Snapshots show the counter-clockwise orbit of the secondary passing behind the slightly tilted disk and then in front of the slightly tilted disk.  The disk appears to change in width as the disk is elliptical; the disk's semi-minor axis is along the stars' line-of-centers.  The flow of the accretion stream above and below the disk is shown by small continuous arrows.  As shown, the accretion stream flows mostly over the disk for nearly $\sim$1/2 orbit and under the disk for $\sim$1/2 orbit.  Because the flow paths are different, a vertical differential pressure generates a lift force as shown by the single large arrow (note, the gravitational force, which is the opposing force, is not shown).  Lift occurs on the near side of the disk, the side closest to the secondary. Lift is in one direction for $\sim$1/2 orbit and in the opposite direction for $\sim$1/2 orbit, resulting in a disk tilt around the line of nodes.  Note, the lift force is not equal in each snapshot, lift occurs on the near side of the disk, and lift does not always occur in the same direction over the orbit.  As a result, we should expect the disk to wobble over time.  The numerical simulation, from which this cartoon was generated, supports the idea of lift as a source to disk tilt in Montgomery \& Martin (2010)

Figure 4 shows the disk wobble over time that is slightly more than twenty orbital periods.  Over the first ten orbital periods (top two panels), the accretion stream flows over the disk rim for $\sim$1/2 orbit and then under the disk for $\sim$1/2 orbit. Over the next ten orbital periods (bottom two panels), the gas stream flows under the disk rim for $\sim$1/2 orbit and then flows over the disk for $\sim$1/2 orbit.  Over time, the disk wobbles backward relative to the direction of orbital motion, and the location on the disk rim where the gas stream transitions from flowing over to under moves in the retrograde direction (an analogy is the constant westward movement of the First Point of Aries along Earth's equator over time as shown in Montgomery 2009b).  As shown, for $\sim$half the retrograde precessional cycle, the disk is tilted towards the secondary and for $\sim$half of the retrograde precession cycle, the disk is tilted away.  

The simulation shows a disk precession that is slightly more than twenty orbital periods.  Even though the disk has layers of different density and shape which affect the disk's moment of inertia (see Montgomery 2009b), we can analytically estimate the disk's retrograde precession period.  If the titled disk is circular, differentially rotating, and of constant density, the retrograde precession period is about 24 orbital periods (see Montgomery 2009b) which is in the ball-park of the simulation.

\section{Impact Effects \& Discussion}
We can perform a simple exercise to determine the amount of energy and force delivered by the gas stream to the collision at the disk face.  We assume a gas stream particle starts at rest infinitely far from the primary star.  From the simulation, we know that the gas stream strikes the disk at the bright spot, a location that is $\sim$0.3$d$.  From total mechanical energy and conservation of energy, we find $v_{i}\sim6\times10^{5}$ ms$^{-1}$ for the speed of the gas stream particle just before impact at the bright spot.  This speed is an order of magnitude more than the speed of fragment A from comet Shoemaker-Levy-9 that struck Jupiter.  As shown in Figure 2, the gas stream flows under the disk at the bright spot and, therefore, the collision at the bright spot is nearly elastic.  From conservation of momentum and kinetic energy, the gas stream particle speed after the collision is nearly the same as the initial speed, assuming equal mass particles in the disk and stream.  With a $\sim$twelve degree trajectory angle as shown in Figure 3, the vertical component of gas stream speed upon impact with the disk face is approximately 20\% the initial speed.  As Figure 2 does not show the gas stream penetrating through the disk after colliding with the disk face, the gas stream must decelerate to zero velocity within the disk.  If we conservatively assume the gas stream penetrates to a depth $0.01d$ after collision with the disk face, then $\sim$70 seconds are needed to uniformly decelerate a gas particle at a rate $\sim1.5\times10^{3}$ ms$^{-2}$.  The energy released to the disk face at the point of impact per gram of infalling gas stream matter is $\sim10^{11}$ ergs (per gram).  If the mass of each gas particle is $m\sim3.5\times10^{14}$ kg (Montgomery, 2009a), then the total energy released to the disk face at the point of impact is $10^{34}$ ergs.  If all this energy from the impact is used in emission from a surface area equal to the gas stream width, then the luminosity is $\sim10^{30}$ ergs s$^{-1}$ which is in the X-ray band.  Like the soft X-rays images of Jupiter during the comet Shoemaker-Levy 9 K-fragment impact (Waite et al., 1995), the impact onto inner disk annuli may be observable.  Like the fireball and vertical plume generated from the comet Shoemaker-Levy 9 impact, a gas particle impact to the disk face may produce a vertically extended reprocessing region (we see yellow crosses in upper left panel of Figure 2 near the impact site but more particles are needed to improve the resolution) such as that suggested for DW UMa (Hoard et al., 2010).  Further, temperatures at this impact site are more than enough to excite strong emission lines such as He II $\lambda$4686 from optically thick, inner disk annuli as suggested by Sing et al. (2007).  As $\sim$one-third of CVs known to have strong He II $\lambda$4686 emission lines are non-magnetic (Szkody et al. 1990), our work suggests that these same systems may have tilted disks.  

\section{Conclusions}
We find that over long periods of (computational) time, flat accretion disks can tilt around the line of nodes in non-magnetic systems with no strong central radiation source.  After several hundreds of orbital periods have passed, we notice that the disk has settled, with the coldest particles falling to the disk midplane, forming a geometrically thin, cold, dense layer.  As the overlying layers are less dense and hotter, they are more pliable and therefore more easily change shape due to impact from the gas stream.  After the disk has tilted, the rim shape looks like an airfoil.  As the accretion stream passes over the disk for nearly half an orbit and under the disk for nearly the last half orbital period, a vertical differential pressure may develop causing a disk tilt around the line of nodes.    

The tilting of a disk by lift force reported here should be tested using a larger number of particles. Although not shown in this work, tilt does occur with disks of different numbers of particles or, similarly, disks of different mass.  However, $\emph{the}$ disk mass that does not yield a disk tilt has not yet been found.  We expect the more massive the disk, the harder to tilt by lift, keeping all other variables unchanged.

Although the simulation shown in this work is a non-magnetic CV systems, the physics of tilted disks shown here applies to all accreting disks including black hole and neutron star systems.  Significantly larger disks, such as protostellar disks, will not have tilted outer annuli as the speed of the infalling gas from rest is subsonic at the rim of the very large disks.  However, as the infalling gas and dust drops further into the potential well of the protostar, the speed increases to supersonic.  Any difference in flow path above and below the inner disk at these supersonic speeds may result in a lift of the inner annuli.  For example, the gas passing through a gap formed in a Class II transitional disks is supersonic.  This combined with a geometrically thinner inner disk may result in a tilt of the inner disk and thus mass transport to inner annuli in short timescales.  This dynamical effect may contribute to planet formation, and is a subject of future paper(s).

\section*{Acknowledgments}
This research was supported in part by NASA through the American Astronomical Society's Small Research Grant Program.  We would like to thank Matt Wood for reading an initial version of this paper and the anonymous reviewer for providing constructive feedback.

\newpage

 %%FIGURE 1 %%%%%%%%%%%%%%%%%%%%%%%%
\begin{figure}
\includegraphics[scale=0.9]{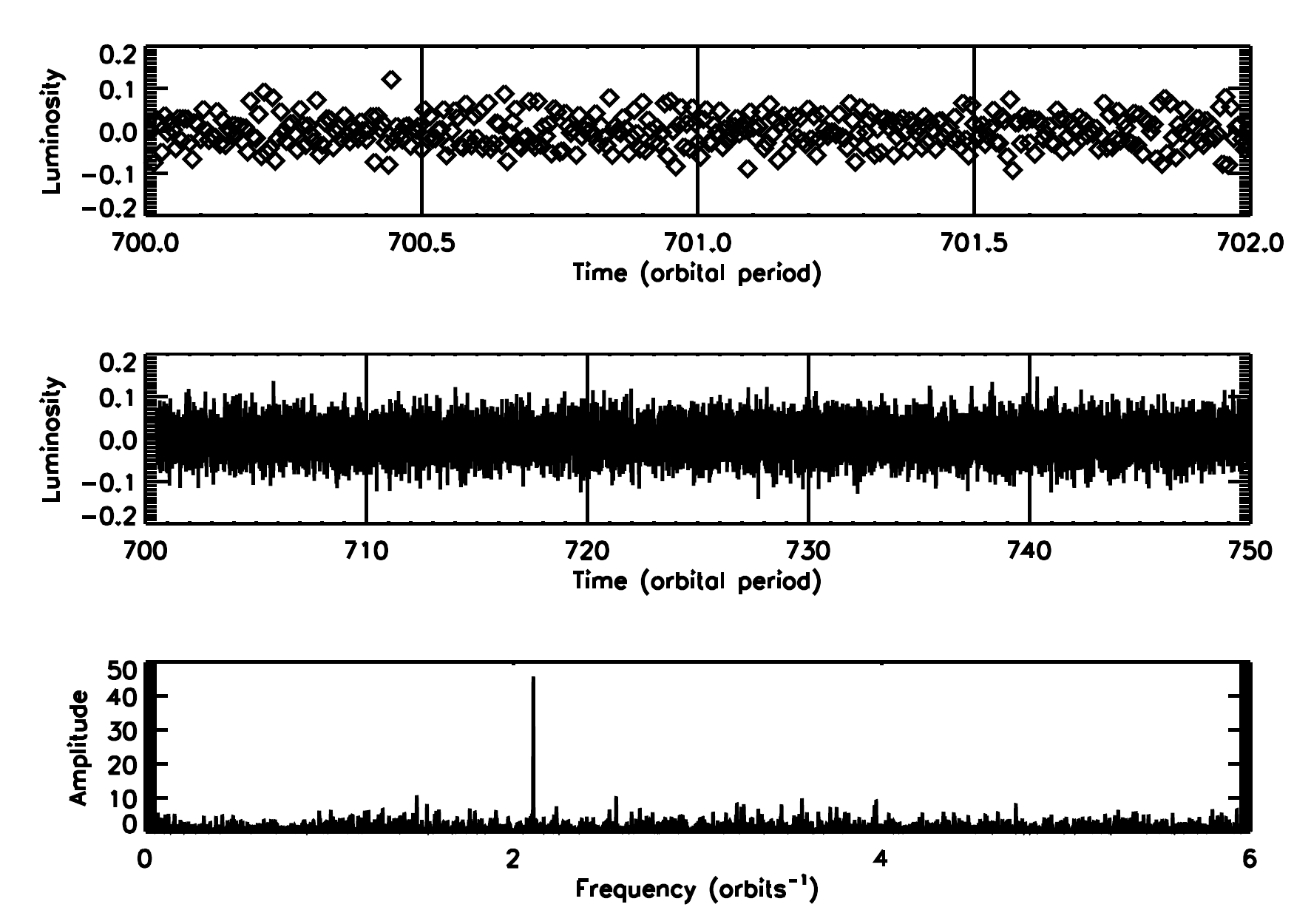}
\caption{Fractional bolometric light curve starting from orbit 700 in the q=0.45 simulation, when the disk begins to tilt around the line of nodes (top two panels), and Fourier transform of orbital periods 700-1000 (lowest panel) are shown.  In the top panel, two hundred data points are sampled per orbit (diamonds).  The top panel only shows the first two orbits and therefore has a different time scale than the middle panel.  Neither the top nor middle panel shows smoothed data.  The negative superhump period $P_{-}$=0.950$P_{orb}$ is seen in the lowest panel.   
}
\label{Figure 1.}
\end{figure}
%%%%%%%%%%%%%%%%%%%%%%%%%%%%%%%%%%%

\newpage
 %%FIGURE 2 %%%%%%%%%%%%%%%%%%%%%%%%
\begin{figure}
\includegraphics[scale=0.9]{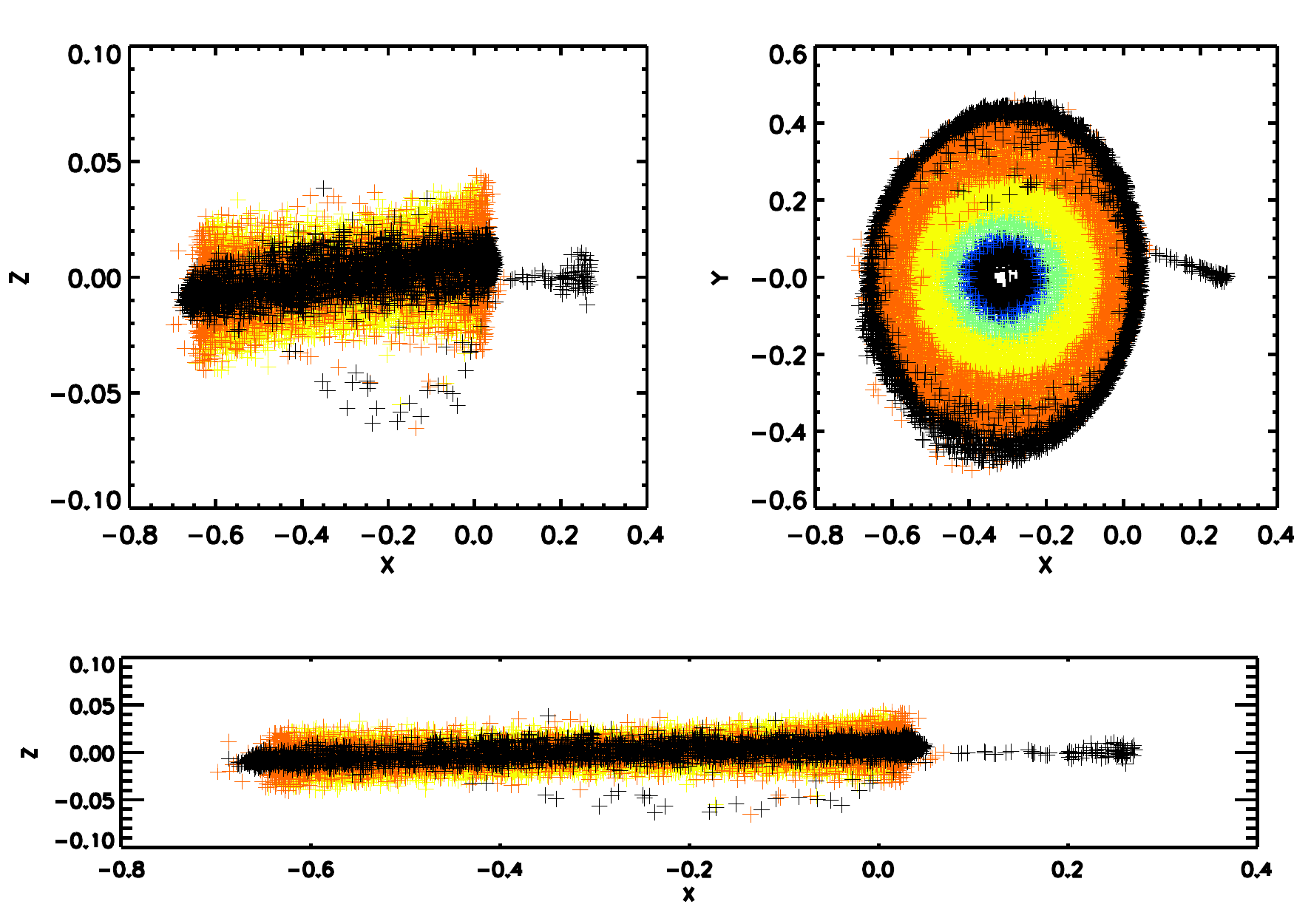}
\caption{Edge-on and face-on snapshots of q=0.45 simulation at time 1230.0$P_{orb}$ in the simulation showing isometric scaling (lower panel, upper right panel) and non-isometric scaling (upper left panel) to emphasize the disk tilt and gas stream paths.  Coolest-to-warmest colors are black, orange, yellow, green, blue, black.  Due to disk tilt, the gas stream strikes the disk at the bright spot, passes under the disk rim, and splashes onto the disk face.  Dislodged bright spot particles also follow a similar path.  Note the coolest bolometric emission in the face-on plot (i.e., black disk particles) are plotted over the warmer bolometric emission to emphasize the location of collision between the gas stream and dislodged particles and the disk face.  
 }
\label{Figure 2.}
\end{figure}
%%%%%%%%%%%%%%%%%%%%%%%%%%%%%%%%%%%

\newpage
 %%FIGURE 3 %%%%%%%%%%%%%%%%%%%%%%%%
\begin{figure}
%\rotatebox{90}{\includegraphics[scale=0.5]{Fig3.pdf}}
\includegraphics[scale=0.9]{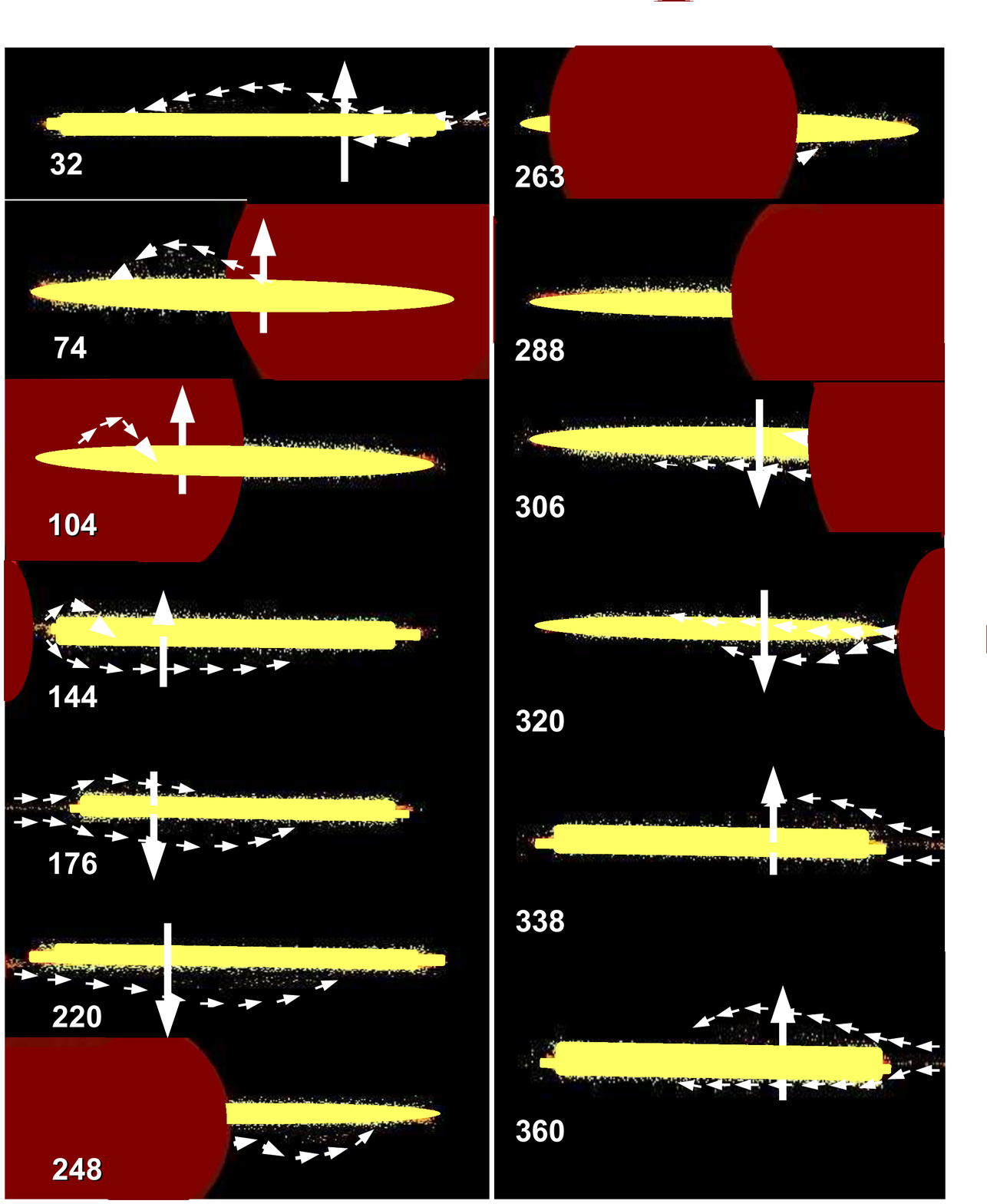}
\caption{Cartoon edge-on snapshots of the q=0.45 simulation are shown, and snapshots are labeled in degrees for one counter-clockwise orbit.  The accretion stream path is shown by small continuous arrows.  As shown, the accretion stream flow path is not the same above and below the disk; the disk has a slight tilt. As a result of the different accretion stream flow paths above and below the disk, lift acts on the disk as shown by the single large arrow.  Lift occurs always on the near side of the disk, the side of the disk nearest to the secondary.   
}
\label{Figure 3.}
\end{figure}
%%%%%%%%%%%%%%%%%%%%%%%%%%%%%%%%%%%

\newpage
 %%FIGURE 4 %%%%%%%%%%%%%%%%%%%%%%%%
\begin{figure}
%\rotatebox{90}{\includegraphics[scale=0.5]{Fig3.pdf}}
\includegraphics[scale=1.4]{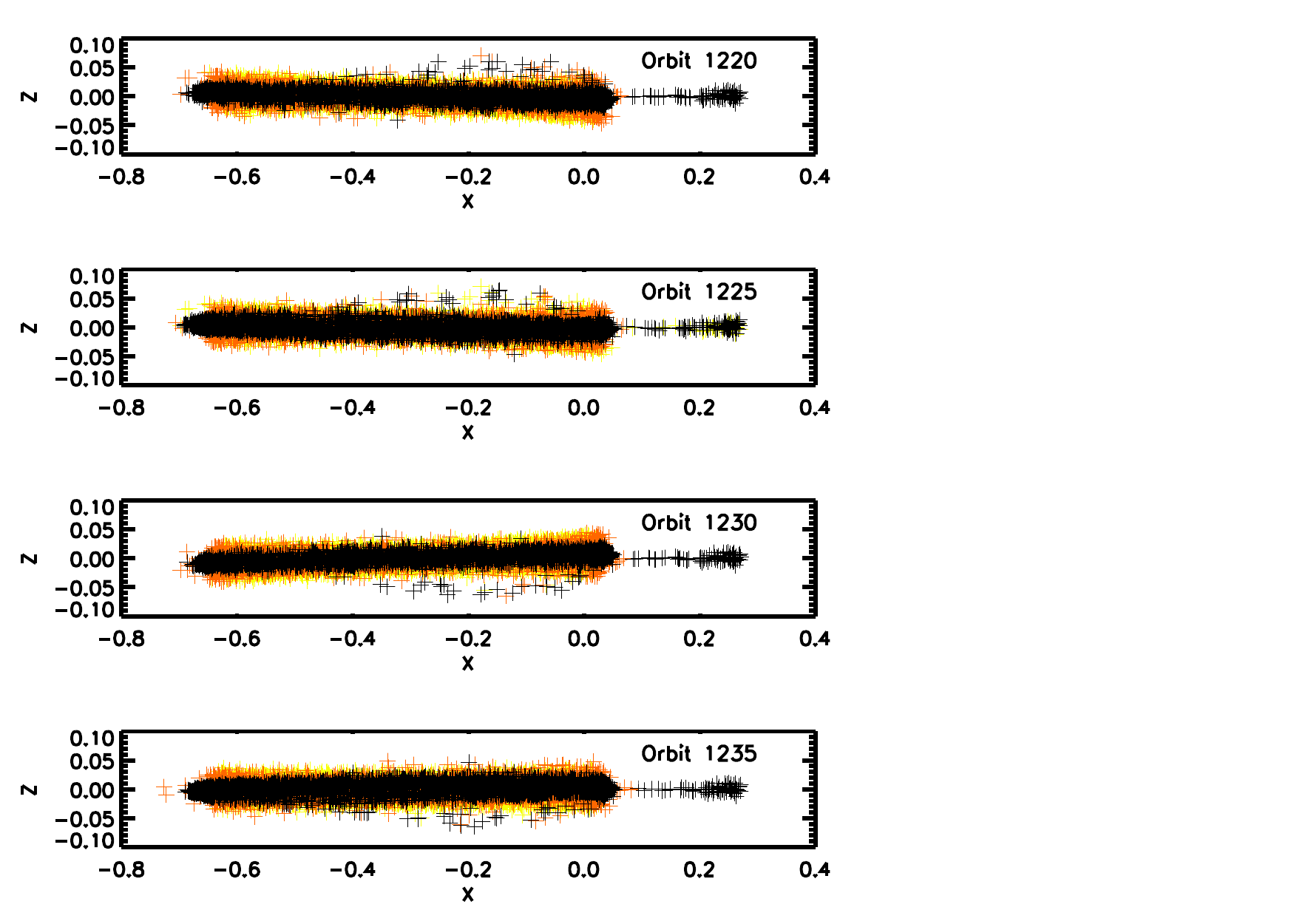}
\caption{Edge-on snapshots of the q=0.45 simulation at various times showing precession.  Coolest-to-warmest colors are black, orange, yellow, green, blue, black.  Due to the disk wobble, orbit 1225 and orbit 1235 show the disk tilted toward you or away from you and hence the disk appears wider that an orbit 1220 or orbit 1230. }
\label{Figure 4.}
\end{figure}
%%%%%%%%%%%%%%%%%%%%%%%%%%%%%%%%%%%

%\begin{thebibliography}{99}
%\bibitem{b1} Barrett P., O'Donoghue D., \& Warner B., 1988, MNRAS, 233, 759 
%\bibitem{b2} Bisikalo D.V., Boyarchuk A.A., Kaygorodov P.V., \& Kuznetsov O.A., 2003, ARep. 47, 809
%\bibitem{b3} Bisikalo D.V., Boyarchuk A.A., Kaygorodov P.V., Kuznetsov O.A., \& Matsuda T., 2004, ARep, 48, 449
%\bibitem{b4} Bonnet-Bidaud J.M., Motch C., \& Mouchet M., 1985, A\&A, 143, 313
%\bibitem{b5} Harvey D., Skillman D.R., Patterson J., \& Ringwald F.A., 1995, PASP, 107, 551
Hoard D.W., Lu T.-N., Knigge C., Homer L., Szkody P., Still M., Long K.S., Dhillon V.S., \& Wachter S., 2010, AJ, 140, 1313 

\section{References}
Barrett P., O'Donoghue D., \& Warner B., 1988, MNRAS, 233, 759 
Bisikalo D.V., Boyarchuk A.A., Kaygorodov P.V., \& Kuznetsov O.A., 2003, ARep. 47, 809
Bisikalo D.V., Boyarchuk A.A., Kaygorodov P.V., Kuznetsov O.A., \& Matsuda T., 2004, ARep, 48, 449
Bonnet-Bidaud J.M., Motch C., \& Mouchet M., 1985, A\&A, 143, 313
Harvey D., Skillman D.R., Patterson J., \& Ringwald F.A., 1995, PASP, 107, 551
Hoard D.W., Lu T.-N., Knigge C., Homer L., Szkody P., Still M., Long K.S., Dhillon V.S., \& Wachter S., 2010, AJ, 140, 1313 
Montgomery M.M. \& Bisikalo D.V., 2010, MNRAS, 405, 1397
Montgomery M.M. \& Martin E.L., 2010, ApJ, 722, 989
Montgomery M.M., 2009a, MNRAS, 394, 1897
Montgomery M.M., 2009b, ApJ, 705, 603
Murray J.R. \& Armitage, P.J. 1998, MNRAS, 300, 561
Murray J.R., Truss M.R., \& Wynn G.A., 2002, in ASP Conf. Proc. 261, The Physics of Cataclysmic Variables and Related objects, ed. B.T. G$\ddot{a}$nsicke, K. Beuermann, \& K. Reinsch (San Francisco, CA: ASP), 416
Patterson J., Thomas G., Skillman D.R., \& Diaz M., 1993, ApJS, 86, 235
Petterson J., 1977, ApJ, 216, 827 
Pringle J.E., 1996, 281, 357
Shakura S.. \& Sunyaev R.A., 1973, A\&A, 149, 135
Simpson J.C., 1995, ApJ, 448, 822
Sing D.K., Green E.M., Howell S.B., Holberg J.B., Lopez-Morales M., Shaw J.S., Schmidt G.D., 2007, A\&A, 474, 951
Szkody P., Garnavich P., Howell S., \& Kii T., 1990, in Proceedings of the 11th North American Workshop on Cataclysmic Variables and Low Mass X-ray Binaries, ed. Christopher W. Mauche (Cambridge University Press), 251
Terquem C. \& Papaloizou J.C.B., 2000, A\&A, 360, 1031
Waite J.H., Jr., Gladstone G.R., Franke K., Lewis W.S., Fabian A.C., Brandt W.N., Na C., Haberl F., Clarke J.T., Hurley K.C., Sommer M., \& Bolton S., 1995, Science, 268, 1598
Wood M.A., Montgomery M.M., \& Simpson J.C., 2000, ApJ, 535, L39
Wood M.A., Still M.D., Howell S.B., Cannizzo J.K., Smale J.P., 2011arXiv1108.3083W
Wood M.A., Thomas D., Simpson J.C., 2009, 398, 2110

\end{document}